\documentclass[twocolumn]{aastex62}
\usepackage{graphics,epsf}
\usepackage{amsmath}                
\usepackage{amsfonts}               
\usepackage{amssymb}                
\usepackage{epsfig}                 
\usepackage{appendix}
\usepackage{graphicx}
\usepackage{float}
\usepackage{color}
\usepackage[para,online,flushleft]{threeparttable}

\newcommand{\cm}{{~\rm cm}}

\newcommand{\km}{{~\rm km}}
\newcommand{\s}{{~\rm s}}

\newcommand{\K}{{~\rm K}}
\newcommand{\erg}{{~\rm erg}}
\newcommand{\yr}{{~\rm yr}}

\newcommand{\kpc}{{~\rm kpc}}

\begin{document}

\title{Jet-driven AGN feedback in galaxy formation before black hole formation}

\email{ ealealbh@gmail.com; soker@physics.technion.ac.il}

\author{Ealeal Bear}
\affiliation{Department of Physics, Technion – Israel Institute of Technology, Haifa 3200003, Israel}

\author{Noam Soker}
\affiliation{Department of Physics, Technion – Israel Institute of Technology, Haifa 3200003, Israel}
\affiliation{Guangdong Technion Israel Institute of Technology, Guangdong Province, Shantou 515069, China}

\begin{abstract}
We propose a scenario where during galaxy formation an active galactic nucleus (AGN) feedback mechanism starts before the formation of a supermassive black hole (SMBH). The supermassive star (SMS) progenitor of the SMBH accretes mass as it grows and launches jets. 
We simulate the evolution of SMSs and show that the escape velocity from their surface is $\approx {\rm several} \times 10^3 \km \s^{-1}$, with large uncertainties. We could not converge with the parameters of the evolutionary numerical code \textsc{MESA} to resolve the uncertainties for SMS evolution. Under the assumption that the jets carry about ten percent of the mass of the SMS, we show that the energy in the jets is a substantial fraction of the binding energy of the gas in the galaxy/bulge. Therefore, the jets that the SMS progenitor of the SMBH launches carry sufficient energy to establish a feedback cycle with the gas in the inner zone of the galaxy/bulge, and hence, set a relation between the total stellar mass and the mass of the SMS. As the SMS collapses to form the SMBH at the center, there is already a relation (correlation) between the newly born SMBH mass and the stellar mass of the galaxy/bulge. During the formation of the SMBH it rapidly accretes mass from the collapsing SMS and launches very energetic jets that might unbind most of the gas in the galaxy/bulge. 
\end{abstract}


\section{Introduction} 
\label{sec:intro}

The formation and evolution of supermassive stars (SMSs) that might collapse to form supermassive black holes (SMBHs) have many open questions (e.g., \citealt{OmukaiPalla2003, Visbaletal2014, Sakurietal2015, Luoetal2018, Woodsetal2017, Tagawaetal2019, Woodsetal2019}). The basic scenario is that primordial gas clouds contract to form supermassive primordial Pop III stars that grow to masses of about $10^4 - 10^6 M_\odot$ by high accretion rates (e.g., \citealt{Agarwaletal2012, Johnsonetal2014, Agarwaletal2016}), and then the SMSs collapse to form BHs that by further accretion grow to SMBHs with masses up to $\approx 10^9 M_\odot$ (e.g., \citealt{Ardanehetal2018, Umedaetal2016}). 
The motivation to study this scenario (e.g., \citealt{Whalenetal2013, Ardanehetal2018, Umedaetal2016, Sakurietal2016, Hirschi2017, Ardanehetal2018, Matsukobaetal2019}) comes from the difficulties for other theoretical scenarios to form SMBHs, masses of $10^9 M_\odot$, at high redshifts of $z \ga 6$ (for details and references see, e.g., \citealt{Glover2016, SmithBromm2019}).  
  
We note that there is a debate whether SMSs can form at all, with some arguments for (e.g., \citealt{Umedaetal2016}; for numerical simulations of a collapse to form SMBH see, e.g., \citealt{Shibataetal2016}) and some against this scenario (e.g.  \citealt{DotanShaviv2012, Yoonetal2015, LatifFerrara2016, CorbettMoranetal2018}). 
For example, one of the unsettled issues is whether fragmentation halts the formation process of SMSs (e.g., \citealt{Brommetal1999, Omukaietal2008, Suazoetal2019}) or not (e.g., \citealt{Begelman2010, CorbettMoranetal2018, Suazoetal2019}). The efficiency of fragmentation may (e.g.  \citealt{Hosokawaetal2013}) or may not (e.g., \citealt{CorbettMoranetal2018}) be connected to the metallicity of the parent cloud. 

In the present study we adopt the collapsing SMS scenario for the formation of SMBHs, and discuss its implications to the feedback mechanism at early times of galaxy formation. \cite{Umedaetal2016} argue that SMSs can form and grow by high mass accretion rates into these almost fully convective stars (e.g., \citealt{Uchidaetal2017, Johnsonetal2012}). \cite{TutukovFedorova2008} claim that at zero metallicity even when one consider the stellar wind, the star can reach a mass of $10^6 M_\odot$.
In our study, we assume that the accretion takes place through an accretion disk that launches jets.
  
Many researchers study jets from massive stars of $M \ga 10 M_\odot$ (e.g., \citealt{Fulleretal1986, Garatti2018, McLeodetal2018}), but the situation is more complicated and uncertain in SMSs.
\cite{LatifSchleicher2016} study a SMS with a mass of $10^5 M_\odot$ and with a luminosity of $ 10^6 L_\odot$ that launches jets with a terminal velocity (the velocity at large distances from the star) of $v_j = 1200 \km \s^{-1}$, and argue that the feedback mechanism due to these jets is significant at this early stage of evolution. 
We adopt the view that SMSs launch jets as they accrete mass at high rates.
We will try to estimate the terminal velocities of such jets and their role in the feedback during galaxy formation. 

In section \ref{sec:MESA} we discuss previous calculations and our radii calculations of SMSs. In section \ref{sec:jets} we discuss the implications of our finding to very early feedback in galaxy formation. Our short summary is in section \ref{sec:summary}. 
  
\section{Supermassive stars with MESA} 
\label{sec:MESA}
There are many numerical and physical difficulties in simulating SMSs, $M \ga 10^4 M_\odot$, with the stellar evolution code \textsc{MESA} (Modules for Experiments in Stellar Astrophysics, version 10398; \citealt{Paxtonetal2011, Paxtonetal2013, Paxtonetal2015, Paxtonetal2018, Paxtonetal2019}). Thorough studies of using \textsc{MESA} exist up to stellar masses of $1000 M_\odot$ (e.g., \citealt{Paxtonetal2011, Smith2014, FullerRo2018}), but only scarce numerical evolution, mostly with other numerical codes, exist for more massive stars. Due to these difficulties, we limit our study only to some general stellar parameters relevant to our goal of exploring very early feedback in galaxy formation. Namely, we study the radius as function of mass for these SMSs, $M_{\rm SMS} \simeq 10^4- 10^6 M_\odot$. 

We limit our study to a metallicity value of $Z=0$, as it is appropriate for SMSs that live at the very early phases of galaxy evolution, before even SMBH are formed. 
{{{{With higher metallicities our proposed scenario encounters difficulties. 
Firstly, a higher metallicity increases the opacity and therefore causes a larger stellar radius that in turn reduces the gravitational energy that the accreted mass releases. This makes the energy of the jets from the SMS smaller. Secondly, for metallicities of $Z \ga 5 \times 10^{-6} Z_\odot$ when dust is included or $Z \ga 3 \times 10^{-4} Z_\odot$ when dust is absent, the cloud fragments to many lower mass stars, rather than one SMS (e.g., \citealt{Omukaietal2008}).  }}}}

{{{{ Because we `stretch' the mass domain of \textsc{MESA} to SMSs, we compare the radii we derive to values from other studies in the literature \citep{Begelman2010, Hosokawaetal2012, Hosokawaetal2013, HaemmerleMeynet2019}. We find that our values fall within the large range of radii that these different studies give, and therefore we believe that the usage of \textsc{MESA} for the present study is adequate. Definitely more development of \textsc{MESA} is required to follow these SMSs. }}}}
   
Earlier studies using different numerical codes and analytical estimates did not reach full agreement on the radii of SMS. 
While some derived radii of $R_{\rm SMS} \simeq 10^4 R_\odot$ for SMS of masses of $M_{\rm SMS} \simeq 10^5 M_\odot$ (e.g., \citealt{Hosokawaetal2013, HaemmerleMeynet2019}), other obtained smaller radii, like \cite{Suraceetal2019} who claim that SMSs are bluer, {{{{surface temperatures of $20,000 - 40,000 \K$, }}}} and have typical radii of only about $ {\rm few} \times 10^3 R_\odot$. Some of the differences might result from different conditions. For example, \cite{Suraceetal2019} neglected the effect of radiation pressure on accreted mass, and \cite{Hosokawaetal2016} included accretion during the evolution. In the simulation by \cite{Hosokawaetal2013}, the stellar radius increases monotonically with the stellar mass as long as the accretion rate stays above $\dot M >10^{-1} M_\odot \yr^{-1}$ (for more details see their figure 6). \cite{Hosokawaetal2013} conclude that pulsational mass-loss and stellar UV feedback do not significantly affect the evolution of SMSs that grow by rapid mass accretion rates. 
As well, \cite{HaemmerleWoodsetal2018} find large variations in the radii when they use different accretion rates onto the SMSs.

With these earlier large uncertainties in mind, we turn to describe our \textsc{MESA} numerical results.  
We describe the setting of \textsc{MESA} in Appendix \ref{sec.Numerical setup}. Readers interested only in the results can skip Appendix \ref{sec.Numerical setup} and continue with the description below. 

We take the common practice in \textsc{MESA} and divide the evolution to pre-main sequence (PMS) and later phases (e.g., \citealt{ShiodeQuataert2014}). To minimize the parameter search and fine tuning, when possible, we use the default settings and numerical parameters of \textsc{MESA}, while in other cases we set different parameters that allow us to follow evolution up to core hydrogen depletion. Numerical difficulties dictate the numerical termination time of the evolution to be when the hydrogen in the core is depleted down to $X=10^{-5}$. 

\begin{figure*}
 \hskip -2.00 cm
 \includegraphics[trim= 0.0cm 0.0cm 0.0cm 0.0cm,clip=true,width=1.25\textwidth]{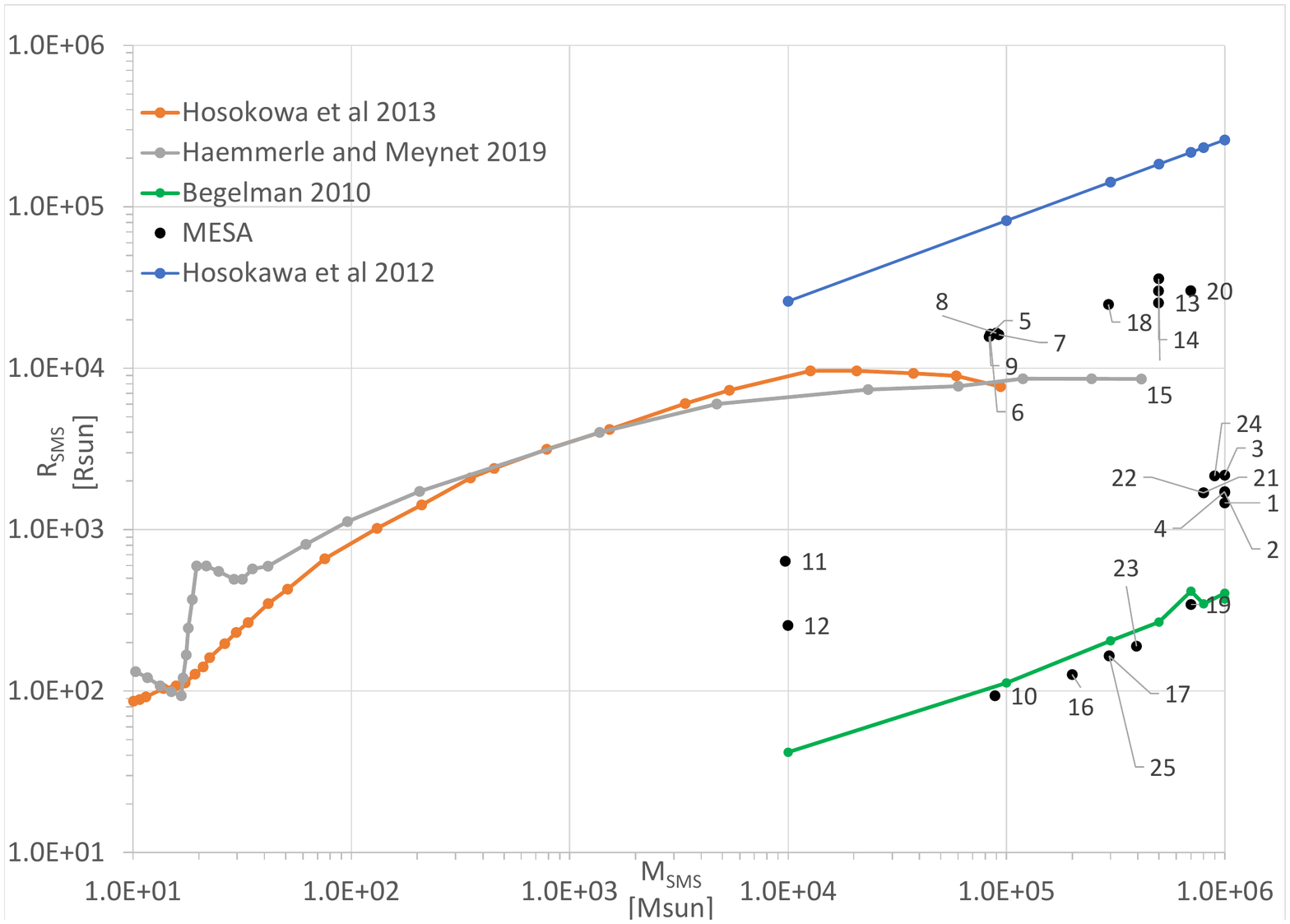}\\
 \vskip -2.00 cm
 \caption{Radius vs. mass. Orange line (and dots) are taken from figure 2 in \cite{HaemmerleMeynet2019}. Gray line (and dots) are taken from figure 2 in \cite{Hosokawaetal2013}. Both were taken at what appears to be zero metallicity. The green line is calculated according to \cite{Begelman2010} relationship for such massive stars. The blue line is an analytical estimate from \cite{Hosokawaetal2012}. Black dots depict our \textsc{MESA} calculations for different setups according to Table 1 in Appendix \ref{sec.Numerical setup}. We show here only runs that converged, as in many cases the numerical code could not convergence.}
 \label{fig:R_vs_M}
 \end{figure*}

In Fig. \ref{fig:R_vs_M} we present the stellar radius at the end of the simulation for each of the 25 simulations whose parameters we describe in Appendix \ref{sec.Numerical setup}. 
 
 We find in our simulations that the radius of a SMS with $M_{\rm SMS} \ga 5 \times 10^5 M_\odot$ fluctuates as the SMS evolves from the main sequence to hydrogen exhaustion, with a typical amplitude of more than $30\%$. We do not study the sources of these fluctuations (numerical or physical) as this requires a thorough study of the numerical code and a comparison between codes, which is beyond the scope of the present study. We rather take the final value at the termination time. 

In Fig. \ref{fig:R_vs_M} we present the final radius as function of the final mass for different runs using \textsc{MESA} (marked by the number of the run). The final mass in our simulations is lower because we include mass-loss. We list the different parameters for each run, including initial and final masses and the stellar age at the end of each simulation, in Table 1 in Appendix \ref{sec.Numerical setup}. 

 The orange line in Fig. \ref{fig:R_vs_M} is from figure 2 of \cite{HaemmerleMeynet2019}, and the gray line is from figure 2 of \cite{Hosokawaetal2013}. The blue line is an analytical estimate from \cite{Hosokawaetal2012}. The green line is based on \cite{Begelman2010} who calculates the stellar radius according to 
\begin{equation}
 R_{\rm SMS} =5.8 \times 10^{13} \left(\frac{\dot{M}_{\rm SMS} t}{10^6 M_\odot}\right)^{0.5} 
\left(\frac{T_c}{10^8 \K}\right)^{-1} \cm,
\label{eq:Begelman}
\end{equation}
where, $\dot{M}_{\rm SMS}$ is the accretion rate at time $t$, and $T_c$ is the core`s temperature. 
In drawing equation (\ref{eq:Begelman}) we take $\dot{M}_{\rm SMS} t$ to be the initial stellar mass and we take $T_c$ from the numerical results of \textsc{MESA} at the end of each simulation. 

\cite{Hosokawaetal2013} study rapidly mass-accreting stars, {{{{ i.e., accretion rates of $\ga 10^{-2} M_\odot \yr^{-1}$, }}}} by numerically solving their interior structure with an energy output (their figure 2) and without an energy output (their figure 7). The difference between these two options for our range of mass ($\ga 10^4 M_\odot$) is negligible and hence we present in our figure the option of the energy output. \cite{Hosokawaetal2013} find (orange line in Fig. \ref{fig:R_vs_M}) that the photosphere of SMSs increases with mass up to $R_{\rm SMS} \simeq 10^4 R_\odot$. Beyond this mass, the radius decreases.

\cite{HaemmerleWoodsetal2018} and \cite{HaemmerleMeynet2019} study the evolution of rotating SMSs up to masses of $\simeq 5 \times 10^5 M_\odot$. \cite{HaemmerleMeynet2019} study the effect of magnetic coupling between the star and its winds on the angular momentum accreted from a Keplerian disc. They find that magnetic coupling can remove angular momentum excess as accretion to the star proceeds. From their figure 2, we learn that the SMS radius increases with SMS mass up to $M_{\rm SMS} \simeq 3 \times 10^4 M_\odot$. From Fig. \ref{fig:R_vs_M} we see that \cite{HaemmerleMeynet2019} and \cite{Hosokawaetal2013} have similar radius to mass relation.
   
In addition to the above numerical results, we also draw two analytical approximations on Fig. \ref{fig:R_vs_M}, one (blue line) by \cite{Hosokawaetal2012} where the radius is proportional to the square root of the mass, and the other (green line) by \cite{Begelman2010} as given in equation (\ref{eq:Begelman}). 
Interestingly, these two lines, more or less, delimit the large range of radii that we find with \textsc{MESA}, and the two similar lines by \cite{HaemmerleMeynet2019} and \cite{Hosokawaetal2013}. 
{{{{ The line by \cite{Hosokawaetal2012} results from Stefan-Boltzmann law combined with that they keep the effective temperature at $T_{\rm eff} \approx 5000 \K$ for these high accretion rates. \cite{Begelman2010} gives the radii of fully convective stars, taking into account changes in the core temperature (for more details see \citealt{Begelman2010} and \citealt{Hosokawaetal2012}). In Appendix \ref{sec.Numerical setup} we show that our models are fully or almost fully convective indeed. While the lower line from \cite{Begelman2010} delimits our lower-radius models quite nicely, the upper line from \cite{Hosokawaetal2012} is about one order of magnitude above the numerical radii, both of earlier studies and of the present study. }}}}

We conclude that we can simulate SMSs with \textsc{MESA}, but we could not converge on the best numerical parameters to use. The reason we can use our present results despite the large uncertainties is that we need only the gravitational potential well of the SMSs. This determines the velocity of the jets that the accretion disk might launch. 
   
\section{The feedback mechanism: energy and jets} 
\label{sec:jets}

There are three phases of feedback from the central object in the scenario we study here. The first one takes place when the SMS grows and launches jets, the second one is a very short phase during which the newly born SMBH launches very powerful jets as it accretes mass at a very high rate from the collapsing SMS, while the third feedback phase might last for up to billions of year (even until present) as the SMBH accretes mass from the interstellar medium. We now examine the first two phases. 

{{{{ First we comment on the launching of jets. Accretion disks around many types of astrophysical objects launch jets (for a review see, e.g., \citealt{Livio1999}). Most relevant to us are disks around young stellar objects and around older main sequence stars that launch jets.  We assume that in a similar way disks around a SMS launch jets. We do not get into the theoretical models for jet launching, but we rather base our assumption on observations. By jets we refer to any bipolar outflow, i.e., two opposite collimated outflows along the polar directions. This outflow can be wide, even with a half-opening angle of $\alpha_j>60^\circ$. For example, \cite{Bollenetal2019} infer that a main sequence companion to an evolved post-asymptotic giant branch (AGB) star launches jets with a half opening angle of $\alpha_j \simeq 76^\circ$ as it accretes mass from the post-AGB star through an accretion disk. Like them, we refer to these as jets and not disk-winds, even that they are very wide.  
}}}}

\subsection{Feedback during SMS growth phase} 
\label{subsec:SMSgrowth}

We expect the SMS to launch jets during its formation, as other young stars do. Namely, the SMS launches jets at the escape velocity and with a mass outflow rate of about $\eta_{\rm SMS,j} \approx 0.1$ of the accretion rate (e.g., \citealt{Livio1999, Pudritzetal2012}).
\cite{Machidaetal2006} find in their simulation of a collapsing primordial cloud of $5.1 \times 10^4 M_\odot$ that a fraction of $3\% -10\%$ of the accreting matter is blown off from the central region. This occurs before the formation of a massive central star. We expect that the presence of a massive central star will increase the fraction of mass that the jets carry (e.g., \citealt{Livio1999}).

From Fig. \ref{fig:R_vs_M} we find the escape velocity for $M_{\rm SMS} \ga 10^4 M_\odot$ to be $v_j =v_{\rm esc} \simeq 6000 \km \s^{-1} (M_{\rm SMS}/10^6 M_\odot)^{1/2}$.
The total energy the jets carry is then 
\begin{equation}
\begin{split}
E_{\rm SMS, j}  & = \int_0^{M_c} \frac {1}{2} v^2_j \eta_{\rm SMS,j} d M \simeq 9 \times 10^{11} 
\\ &
\times 
\left(\frac{\eta_{\rm SMS,j}}{0.1} \right)
\left(\frac{{M}_{\rm c}}{10^6 M_\odot}\right)^{2}  M_\odot \km^2 \s^{-2} \end{split}
\label{eq:Esmsj}
\end{equation}
where $M_c$ is the SMS mass when it starts collapsing, and $9\times 10^{11} M_\odot \km^2 \s^{-2} =1.8 \times 10^{55} \erg$.
{{{{{ Earlier studies of jets from primordial SMSs include the magneto-hydrodynamical simulations that \cite{Machidaetal2006} conducted. They simulated the launching of jets by a cloud of $5.1 \times 10^4 M_\odot$ that collapses to form a SMS at the early universe.  }}}}} {{{{{ Due to the high angular momentum of the gas that the SMS accretes, very close to the SMS the gas flows inward mainly near the equatorial plane. This leaves the polar directions near the SMS relatively free for the expansion of the jets. This inflow-outflow structure is similar to young stellar objects where observations show jets to expand to large distances. }}}}}

Let us consider the gas in the bulge of such a new galaxy.  
The velocity dispersion of host bulges of SMBHs with masses of $M_{\rm BH} \simeq 10^6 M_\odot$  is $\sigma_e \simeq 60 \km \s^{-1}$  (e.g., \citealt{Gebhardtetal2000, deNicolaetal2019}). As well, the stellar bulges mass is $M_{\rm bulge} \approx 10^3 M_{\rm BH}$ (e.g., \citealt{Schutteetal2019}), where by bulge we refer to the spherical (more or less) component of the galaxy, which might be the entire stellar mass in elliptical galaxies. {{{{ We note that the bulge-to-BH mass ratio depends on late-type versus early-type host galaxies \citep{Davisetal2018, Davisetal2019, Sahuetal2019}, but taking it into account is beyond the scope of this paper. }}}}   We assume a similar gas mass at the early phase of galaxy formation. {{{{ As we have no observations of gas mass at this very early phases of galaxy evolution, this is a strong assumption. We make this assumption from the lack of a better way to proceed. }}}}  
The approximate total virial energy of the gas is therefore $E_{\rm bulge,gas} \approx M_{\rm bulge} \sigma^2_e$, which amounts to   
\begin{equation}
E_{\rm bulge,gas} 
\approx 3.6 \times 10^{12} 
\left(\frac{{M}_{\rm c}}{10^6 M_\odot}\right)  M_\odot \km^2 \s^{-2}  .
\label{eq:Ebulgegas}
\end{equation}

The conclusion from equations (\ref{eq:Esmsj}) and (\ref{eq:Ebulgegas}) is that the energy in the jets during the SMS growth is non-negligible. This is particularly the case if we consider that the jets interact only with the polar gas they encounter along their propagation direction. For a half opening angle of $\alpha_j \la 40^\circ$ the energy in the jets is larger than the gas energy, so the jets can substantially heat and/or unbind some of the gas. 
If the SMS radius is $\approx 10^3 R_\odot$ instead of $\approx 10^4 R_\odot$, then the jets carry about ten times more energy and the feedback with the galaxy formation is much more important during this phase. 

{{{{ As the jet's material interacts with the gas in the bulge it passes through a shock wave that heats the gas to temperatures of $T > 10^{8} \K$. The post-shock density is very low, $n \la 1 \cm^{-3}$, where $n$ is the total particle number density. For example, a mass of $10^5 M_\odot$ that is shocked within $0.1 \kpc$ forms a bubble with a total particle number density of $n=1.7 \cm^{-3}$. The radiative cooling time of the gas in the hot bubble is $\tau_{\rm rad} \simeq 10^{8} \yr$. This time scale is much longer than the life time of SMSs, and the bubble does not have time to lose energy to radiation. The high-pressure bubbles (one bubble for each jet) accelerate the gas. The jets penetrate through the gas by interacting with only a fraction of the bulge-gas along the polar directions. If the jets are not too wide, they might proceed at  $\approx {\rm few} - 10 \%$ of their speed as they inflate the bubbles, i.e., at a velocity of $v_{\rm bub}  \approx 200 - 1000 \km \s^{-1}$. The jets influence the gas at wider angles by shocks and by mixing. The exact jets' propagation velocity through the gas is of less importance than the overall energy budget that we mentioned above.  }}}} 

The life time of a SMS with a mass of $\approx 10^6 M_\odot$ and a radius of $\approx 10^4 R_\odot$ is $t_{\rm SMS} \approx 10^6 \yr$ (see Appendix \ref{sec.Numerical setup}). During that time the jet-inflated bubbles (namely, the jets) can reach a distance of $R_{\rm FB} \approx 0.5 \kpc$, {{{{ for $v_{\rm bub} \approx 500 \km \s^{-1}$.}}}} {{{{Typical bulge sizes are $\approx 0.3-3 \kpc$ (e.g., \citealt{MendezAbreuetal2018}), and somewhat smaller at high red-shifts and low masses bulges (e.g., \citealt{Bruceetaletal2012}). }}}} 
We find that the distances that the jets and the bubbles they inflate propagate is {{{{ a substantial fraction of the bulge sizes. }}}} 
     
The region from which gas can feed the SMS is much smaller. Taking the velocity for the inflow to be as the velocity  dispersion, {{{{ which is about the free fall velocity, }}}} the feeding zone has a radius of $R_{\rm feed} \approx \sigma_e t_{\rm SMS} \approx 0.06 \kpc$.
This radius is about one tenth of the size of the bulge. This is like the situation in cooling flows in clusters of galaxies, where the hot gas extends to hundreds of \kpc, while the gas with a cooling time shorter than the age of the galaxy resides in a region that is only about one tenth of that radius. Namely, the region where feedback takes place is much smaller than the total extent of the gas. It is quite possible that during the very early time of galaxy formation there is a small scale cooling flow \citep{Soker2010}. Here, we strengthen the preliminary idea of \cite{Soker2016Rev} that the feedback with the inner region of the interstellar gas starts during the growth phase of the SMS.

\subsection{Outcomes of SMBH formation} 
\label{subsec:SMBHformation}

Earlier studies examined the process where the newly born SMBH launches jets as it accretes mass from the collapsing SMS (e.g., \citealt{Barkov2010, Matsumotoetal2016, Uchidaetal2017}), and the feedback with the environment that the jets might induce (e.g., \citealt{Matsumotoetal2015}). 
\cite{Matsumotoetal2015} and \cite{Matsumotoetal2016} who discussed gamma ray bursts from these jets took the jet energy to be $E_{\rm BH,j}=\eta_{\rm BH,j} M_c c^2 \approx 10^{55-56}\erg$ with $\eta_{\rm BH,j}=6.2 \times 10^{-4}$, and where $M_c$ is the collapsing mass. \cite{Matsumotoetal2015} estimated the event rate to be about one per year. 
We note that the $M_c=10^5 M_\odot$ SMS model of \cite{Matsumotoetal2015} has a radius of about $10^3 R_\odot$, about one order of magnitude smaller than what we use here. Using a smaller radius would imply much higher jets' velocity, because we assume that the jets' velocity is about equal to the escape velocity from the SMS (section \ref{subsec:SMSgrowth}), and therefore a much more efficient feedback during the growth phase of the SMS. These papers, among others, derive many properties, like gamma ray bursts, that we do not consider here. Our only goal is to argue that AGN feedback during galaxy formation has started before the formation of the SMBH. In section \ref{sec:summary} we will discuss the implications of this very early feedback process.  

Let us derive the basic parameters of the jets for the case we study here, and mention their relevance to the present study. The accretion phase will last for about the free fall time of the outer region of the SMS, {{{{ much as the accretion phase onto the newly born neutron star in core collapse supernovae lasts for about the free fall time of the accreted mass from the inner part of the core (as only the inner part of the core collapses in that case). }}}}  This time is $\tau_{\rm ff} = 0.056 (R/10^4 R_\odot)^{3/2}(M_{\rm c}/10^6 M_\odot)^{-1/2} \yr$.
\cite{Sunetal2017} conducted magnetohydrodynamic simulations of a collapsing SMS and estimated the accreation torus lifetime to be $t \approx 0.002 ({M_{\rm c}}/{10^6}) \yr$, which is  shorter than the free fall time we take here, or about the same if the radius of the SMS is about $10^3 R_\odot$ and not about $10^4 R_\odot$. This very short accretion phase implies that there is no time to establish a feedback cycle with the interstellar medium. As the paper cited above already noticed, we have here an explosion, more similar to low mass long gamma ray bursts. 

However, the total energy in the jets is very large
\begin{equation}
\begin{split}
E_{\rm BH, j}  & = \eta_{\rm BH,j} M_{\rm c} c^2 = 9 \times 10^{13} 
\\ &
\times 
\left(\frac{\eta_{\rm BH,j}}{0.001} \right)
\left(\frac{{M}_{\rm c}}{10^6 M_\odot}\right)  M_\odot \km^2 \s^{-2}. 
\end{split}
\label{eq:EBHj}
\end{equation}
This energy, of $E_{\rm BH, j} = 2 \times 10^{57} \erg$ for the above scaling, can be more than an order of magnitude larger than the binding energy of the gas in the bulge hosting the newly born SMBH (eq. \ref{eq:Ebulgegas}). 
As \cite{Matsumotoetal2015} already discussed, these jets can expel all the gas from the young low-mass galaxy (or bulge). This leaves the galactic stellar mass to be that of the mass of the stars that have been already formed, and the SMBH mass to be about equal to that of the SMS. Namely, {{{{ if our assumption that the SMS progenitor of the SMBH launches sufficiently energetic jets that interact with the gas in the bulge to set a negative feedback cycle is correct, }}}} then this feedback during the SMS growth phase determines the relation between the bulge and SMBH masses. We discuss this further in section \ref{sec:summary}. 

\section{Discussion and Summary} 
\label{sec:summary}

The goal of the present study is to strengthen the case for a very early feedback process between the growth of the total stellar population mass of a bulge (or a galaxy) and the growth of its central massive body. The motivation to consider a very early feedback comes from the correlation between the mass of the SMBH and the properties of its host galaxy (bulge; e.g., \citealt{Benedettoetal2013, GrahamScott2013, Saxtonetal2014}), in particular the total stellar mass. The properties of the bulge-SMBH masses correlation itself shows that this correlation cannot be driven by many mergers of low mass galaxies \citep{Ginatetal2016}. {{{{ Specifically, the merger-only explanation for the correlation predicts that the relative scatter around the mean proportionality relation between the SMBH and bulge masses increases with the square root of the masses. \cite{Ginatetal2016} examined a sample of 103 galaxies and found that the intrinsic scatter increases more rapidly than expected from the merger-only scenario.  }}}} That merger cannot work  hints that there is a very early process, most likely a feedback process, that starts to establish the correlation. {{{{ We assume that the feedback is between the central object by the jets that it launches and the gas in the bulge/galaxy. We note though, that there are scenarios where the stars in the bulge/galaxy, rather than the gas, interact with the central SMBH to determine the correlation (e.g., \citealt{MichaelyHamilton2020}). }}}}

{{{{ There are no direct observations that show that a feedback cycle driven by the interaction of the jets (or winds) that the central star launches and the gas in the bulge/galaxy determines the SMBH-bulge correlation. Indirect supports exist, in particular showing that in some cases the collimated outflows from AGNs are sufficiently massive and energetic to provide a feedback (e.g., \citealt{Borguetetal2013}). 
Direct observations for a feedback exist in cooling flow clusters and galaxies where we see large jet-inflated bubbles that heat the intracluster medium (e.g., reviews by \citealt{Soker2016Rev, Werneretal2019}). But the direct observational supports for cooling-heating feedback cycle of the intracluster medium do not directly show that this sets the correlation between the stellar and SMBH masses. The idea of such a feedback, nonetheless, is quite popular as an explanation for the correlation (e.g. \citealt{Graham2016, Aravetal2020, Terrazasetal2020}), despite that different studies propose different feedback modes of operation (e.g., \citealt{SokerMeiron2011, Beltramonteetal2019}). }}}}
    
We therefore examine the possibility that jets from SMS progenitors of SMBHs have enough energy to start a feedback process even before a SMBH is formed. The jets that these SMSs launch are non-relativistic, but rather have a velocity of $v_j \simeq {\rm several} \times 1000 \km \s^{-1}$ (section \ref{subsec:SMSgrowth}). This is not a problem as observations show that even in evolved (old) AGN non-relativistic outflows of similar velocities can drive a feedback process (e.g., \citealt{Chamberlainetal2015, XuAravetal2019}).  

We summarize our main results first, and then their implications. 
(1) We could not resolve the question of whether SMSs of $\approx 10^6 M_\odot$ have a radius of $R \approx 10^3 R_\odot$ or $R \approx 10^4 R_\odot$. We therefore adopted the results of \cite{HaemmerleMeynet2019} and  \cite{Hosokawaetal2013} and took $R \approx 10^4 R_\odot$ (see recent review by \citealt{Hosokawa2019}). To resolve this question, there is a need for a thorough and detailed study with \textsc{MESA}. These SMSs live for $t_{\rm SMS} \approx 10^6 \yr$. 

(2) Under this assumption and the assumptions that SMSs launch jets at the escape velocity from their surface and that about ten percent of the accreted mass is carried out by the jets, the total kinetic energy of the jets is 
$E_{\rm SMSs,j} \approx {\rm few} \times 10^{55} \erg \simeq 10^{12} M_\odot \km^2 \s^{-2} $ (equation \ref{eq:Esmsj}). This might be a substantial fraction of the energy of the gas in the interstellar medium under the assumption that the gas mass is $M_{\rm bulge} \approx 10^3 M_{\rm BH}$ (equation \ref{eq:Ebulgegas}). 

(3) As we discussed in section \ref{subsec:SMSgrowth}, during the SMS growth phase that lasts for about $10^6 \yr$ (e.g., \citealt{Hosokawaetal2013} and Table 1 in Appendix \ref{sec.Numerical setup}), the jets can propagate through a large fraction of the bulge size (or small new galaxy size), and the jets might establish a feedback with the interstellar medium within about $0.05 \kpc$. A situation similar to that in cooling flows in clusters of galaxies might take place here, but within a much smaller region \citep{Soker2010}. 

(4) As the SMS collapses to form a SMBH, the SMBH accretes mass at a very high rate from the collapsing SMS and is likely to launch jets. As earlier studies concluded, e.g., \cite{Matsumotoetal2015}, the energy in these jets (equation \ref{eq:EBHj}) might be much larger than the binding energy of the interstellar such that the jets expel most or all of this gas. \cite{Klameretal2004}, on the other hand, discuss jet-induced star formation scenario from observation and noted its implication to galaxy formation.

One of the conclusions from our findings is that even in systems where the accretion of gas onto the bulge and the accretion of mass onto the SMBH are negligible after SMBH formation, the correlation between the SMBH mass and total stellar bulge (or dwarf galaxy) mass has been already established at earlier times. Namely, this correlation holds even in the least massive systems. 
\cite{Yang2019}, for example, argue that the correlated growth of the masses of the bulge and of the SMBH it hosts started very early in the Universe, at a redshift of $z=3$ or earlier even. We suggest that this correlated growth started with the formation of the SMS progenitor of the SMBH.

\acknowledgments
{{{{ We thank an anonymous referee for detailed comments that substantially improved the presentation of our suggested scenario. }}}} This research was supported by a grant from the Israel Science Foundation.


\appendix\section{Numerical setup}
\label{sec.Numerical setup}
    
This appendix contains a technical description of the relevant physical and numerical parameters of the numerical code \textsc{MESA}. Parameters that we do not  address here are as in the 'controls' and 'star jobs' default inlists of \textsc{MESA}. 

We could not achieve convergence of SMS models with the default \textsc{MESA} parameters (both regular and those adopted for high mass). Therefore, we changed a few parameters from different categories as we describe below.
{{{{ Our approach was to make minimum modifications to the default parameters of \textsc{MESA}. We started by demanding each simulation to converge and to allow us to determine the stellar radius at the time of hydrogen depletion.
It turned out that whenever we reached a convergence, the stellar radius was within the range of stellar radii that earlier studies in the literature give for SMSs with similar initial masses (Fig. \ref{fig:R_vs_M}). Nonetheless, we encourage future studies to better determine the best \textsc{MESA} numerical parameters for SMSs. 
 }}}}

\subsection{Instabilities}
\label{subsecAinstabilities}

To deal with stellar instability we changed a set of parameters that includes the following. The shear instability parameter $D\_{DSI}$, the Solberg-Hoiland parameter $D\_{SH}$, the secular shear instability $D\_{SSI}$, the Eddington-Sweet circulation parameter $D\_{ES}$, the Goldreich-Schubert-Fricke $D\_{GSF}$ parameter, and the Spruit-Tayler dynamo parameter $D\_{ST}$ (for more details see \citealt{Hegeretal2000, Hegeretal2005}). At default settings, all these instability parameters in \textsc{MESA} are set to 0. Two other parameters: $am\_D\_mix\_factor$ and $am\_nu\_visc\_factor$ (which are part of the algebraic formula that \textsc{MESA} uses to calculate the diffusion coefficient for mixing of material) are $0$ and $1$ in the default setting, respectively. 

We checked the impact of these instability parameters according to \cite{ShiodeQuataert2014} and \cite{Gilkis2018} {{{{ as they supply the full \textsc{MESA}-inlist that we can folow. }}}} \cite{ShiodeQuataert2014} studied massive stars of $M>30M_\odot$ and took $ D\_SSI\_factor=  D\_ES\_factor= D\_GSF\_factor=1.16$, keeping all other parameters at their default values. \cite{Gilkis2018} set the parameters at:
    $am\_nu\_visc\_factor = 0$,
    $am\_D\_mix\_factor = \frac{1}{30}$ \citep{Hegeretal2000},
    and
    $D\_DSI\_factor = D\_SH\_factor =  D\_SSI\_factor =  D\_ES\_factor = D~GSF~factor = 1 $. We also take $D\_ST\_factor = 1$.
    
In our simulations we adopted one of three options. (A) Instability parameters according to \cite{Gilkis2018} and $D\_ST\_factor = 1$, (B) instability parameters according to the \textsc{MESA} default values, and (C) instability parameters according to \cite{ShiodeQuataert2014}. 
We list the parameters, both instability parameters and others, of 25 of our simulations in Table \ref{tab:my-table}. In many other simulations \textsc{MESA} did not converge and we do not present these cases here. 
 
\begin{table}[]
\centering
\begin{tabular}{|l|l|l|l|l|l|l|l|l|l|l|l|l|l|}
\hline
RUN & $M_i$ & $M_f$ & $R_f$ & Age & \multicolumn{3}{l|}{Instabilities} & \multicolumn{3}{l|}{MLT option} & \multicolumn{2}{l|}{EOS} &  \\ \hline
 & [$M_\odot$] & [$M_\odot$] & [$R_\odot$] & [\yr] & A & B & C & I & II & III & a & b & \begin{tabular}[c]{@{}l@{}}Max\\   model\end{tabular} \\ \hline
1 & 1.00E+06 & 1.00E+06 & 1.47E+03 & 9.34E+04 & V &  &  & V &  &  &  & V & 1d5 \\ \hline
2 & 1.00E+06 & 1.00E+06 & 1.70E+03 & 1.49E+05 & V &  &  &  & V &  &  & V & ! \\ \hline
3 & 1.00E+06 & 1.00E+06 & 2.17E+03 & 2.35E+05 &  & V &  & V &  &  &  & V & ! \\ \hline
4 & 1.00E+06 & 1.00E+06 & 1.72E+03 & 2.21E+05 & V &  &  & V &  &  & V & V & 1d5 \\ \hline
24 & 9.00E+05 & 8.99E+05 & 2.15E+03 & 3.35E+05 & V &  &  & V &  &  & V & V & 5d5 \\ \hline
21 & 8.00E+05 & 8.00E+05 & 1.69E+03 & 1.58E+05 & V &  &  & V &  &  & V & V & 1d5 \\ \hline
22 & 8.00E+05 & 8.00E+05 & 1.69E+03 & 1.58E+05 & V &  &  & V &  &  & V & V & 5d5 \\ \hline
19 & 7.00E+05 & 7.00E+05 & 3.44E+02 & 3.07E+04 & V &  &  & V &  & V &  & V & 1d5 \\ \hline
20 & 7.00E+05 & 6.98E+05 & 3.03E+04 & 1.19E+06 & V &  &  & V &  &  &  & V & 1d5 \\ \hline
13 & 5.00E+05 & 4.97E+05 & 2.54E+04 & 1.58E+06 & V &  &  &  & V &  &  & V & 1d5 \\ \hline
14 & 5.00E+05 & 4.98E+05 & 3.01E+04 & 1.21E+06 & V &  &  & V &  &  & V & V & 1d5 \\ \hline
15 & 5.00E+05 & 4.98E+05 & 3.59E+04 & 1.03E+06 &  & V &  & V &  &  & V & V & 1d5 \\ \hline
23 & 4.00E+05 & 3.94E+05 & 1.89E+02 & 4.41E+04 & V &  &  & V &  & V &  & V & 5d5 \\ \hline
25 & 3.00E+05 & 2.96E+05 & 1.65E+02 & 4.85E+04 & V &  &  & V &  & V &  & V & 1d5 \\ \hline
17 & 3.00E+05 & 2.96E+05 & 1.65E+02 & 4.85E+04 & V &  &  & V &  & V &  & V & ! \\ \hline
18 & 3.00E+05 & 2.93E+05 & 2.48E+04 & 1.61E+06 & V &  &  & V &  &  & V & V & 5d5 \\ \hline
16 & 2.00E+05 & 2.00E+05 & 1.27E+02 & 5.69E+04 & V &  &  & V &  & V &  & V & ! \\ \hline
5 & 1.00E+05 & 8.43E+04 & 1.63E+04 & 1.58E+06 & V &  &  &  & V &  &  & V & ! \\ \hline
6 & 1.00E+05 & 8.41E+04 & 1.59E+04 & 1.58E+06 & V &  &  & V &  &  & V & V & ! \\ \hline
7 & 1.00E+05 & 9.24E+04 & 1.61E+04 & 1.55E+06 &  &  & V & V &  &  &  & V & ! \\ \hline
8 & 1.00E+05 & 8.99E+04 & 1.65E+04 & 1.57E+06 &  & V &  & V &  &  &  & V & ! \\ \hline
9 & 1.00E+05 & 8.33E+04 & 1.57E+04 & 1.58E+06 & V &  &  & V &  &  &  & V & ! \\ \hline
10 & 1.00E+05 & 8.86E+04 & 9.35E+01 & 7.80E+04 & V &  &  & V &  & V &  & V & ! \\ \hline
11 & 1.00E+04 & 9.72E+03 & 6.38E+02 & 1.54E+06 & V &  &  & V &  &  &  & V & ! \\ \hline
12 & 1.00E+04 & 9.98E+03 & 2.55E+02 & 1.52E+06 &  & V &  & V &  &  &  & V & ! \\ \hline
\end{tabular}
\caption{The stellar evolution simulations that converged, i.e., could reach the time of core hydrogen depletion, by descending initial mass. All are marked on Fig. \ref{fig:R_vs_M}. 
The first five columns give the run number, the initial mass $M_i$, the final mass $M_f$ which is somewhat lower due to a wind, the final stellar radius $R_f$, and the age at the termination of the calculation. The other columns list the \textsc{MESA} parameters as we describe in the text. 
 The first group of three columns refers to the the instabilities options (section \ref{subsecAinstabilities}), A: as in  \cite{Gilkis2018} and $D\_ST\_factor = 1$, (B) According  to the \textsc{MESA} default values, and (C) according to \cite{ShiodeQuataert2014}. 
The first two MLT columns are (I) Henyey for PMS and Cox for main sequence and later, (II) Henyey for both PMS and later evolution (section \ref{subsecAmixingLength}). The last option (III) allows the code to boost efficiency of energy transport (\textsc{MESA} parameter, $okay\_to\_reduce\_gradT\_excess$).
In the EOS columns `a' and `b' represent activating the options $Include\_dmu\_in\_eps\_grav$ and $fix\_eps\_grav\_transition\_to\_grid$,  respectively (for details see \textsc{MESA} defaults).
The last column: Max model, limits the number of models in the code. 
The mark `\rm{!}' in the last column of Table \ref{tab:my-table} means that we set no limit on the number of models, i.e., we worked with the default options.}
\label{tab:my-table}
\end{table}


\subsection{Structure parameters}
\label{subsecAstructure}

We activated two parameters ($include\_dmu\_dt\_in\_eps\_grav$ {{{{e.g., \citealt{Fulleretal2015} and \citealt{KissinThompson2018}}}}} and $fix\_eps\_grav\_transition\_to\_grid$) that are deactivated in the default setting.
The parameter $include\_dmu\_dt\_in\_eps\_grav$ includes the contribution from composition changes when activated. We activated this parameter in many of our simulations as recommended for high temperature in \textsc{MESA}.
The purpose of $fix\_eps\_grav\_transition\_to\_grid$ is to fix the transition region of the mesh and it helps with convergence near the Eddington limit {{{{ (as in some of the simulation of \cite{Paxtonetal2018}; their Figs. 49 - 51). }}}} 
We note that run 22 was done twice. Once as noted in the table and the second time when $use\_ODE\_var\_eqn\_pairing$ is activated (default setting is deactivated). The final radius and age were identical.
 
\subsection{Mixing parameters}
\label{subsecAmixing}

The parameter $okay\_to\_reduce\_gradT\_excess$ adjusts the gradient of the temperature to boost efficiency of energy transport. We activated this parameter following \cite{Fulleretal2015} and \cite{Quataertetal2016}  (in the default setting it is deactivated). We note that \cite{Keszthelyietal2017} also activated it for a stars of $15M_\odot$.

\subsection{Mixing length theory (MLT)}
\label{subsecAmixingLength}

We checked two settings for the $MLT~option$, Cox \citep{CoxGiuli1968} which is the default setting and Henyey \citep{Henyeyetal1965}. The Henyey setting allows the convective efficiency to vary with the opaqueness of the convective element. Near the outer layers of stars, this is an important effect for convective zones, while the Cox setting assumes high optical depths and no radiative losses (for more details see \citealt{Paxtonetal2011}).
We achieved convergence for several combinations as indicated in Table \ref{tab:my-table}. {{{{ We consulted the blog of \textsc{MESA} users in choosing these combinations. \footnote{https://lists.MESAstar.org/pipermail/MESA-users/} }}}} 

\begin{figure*}
 \hskip -2.00 cm
 \includegraphics[trim= 0.0cm 0.0cm 0.0cm 0.0cm,clip=true,width=1.25\textwidth]{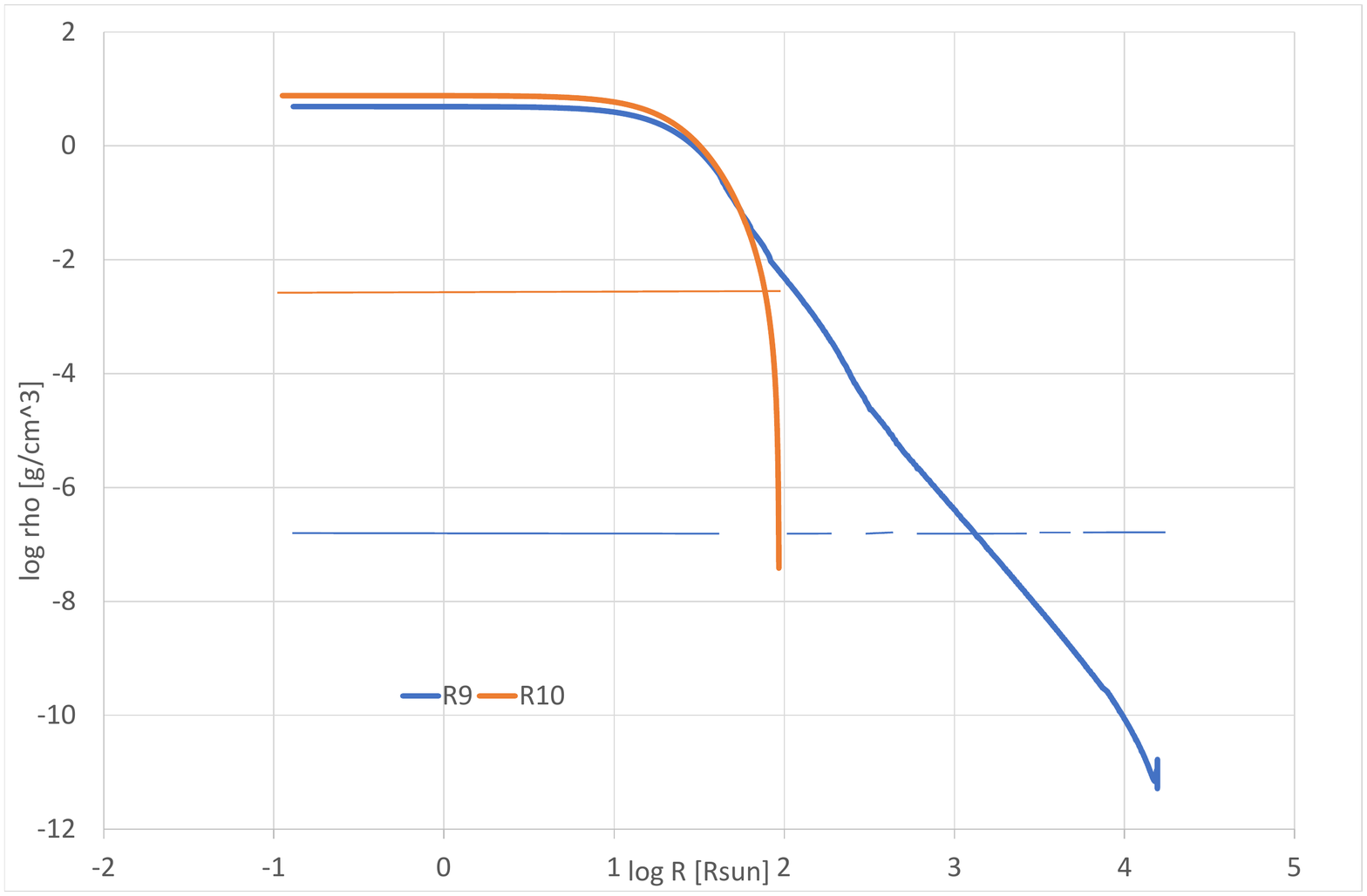}\\
 \vskip -2.00 cm
 \caption{ The density versus radius ($\log \rho $ vs. $ \log R$) at the termination time of the simulation, i.e., hydrogen depletion in the core, for two models with the same initial mass (before mass-loss) of $10^5 M_\odot$. The difference in the input between the R9 model (blue line) and the R10 model (orange line) is that in R10 we boost the efficiency of energy transport (Table \ref{tab:my-table}), which in turn leads to a smaller radius. The horizontal lines with the respective colors depict the convection zones.}
 \label{fig:rho_vs_R_R9_10}
 \end{figure*}

{{{{ In some models we allowed the code to boost efficiency of energy transport (we mark this by `V' in option~III in the table; \textsc{MESA} parameter, $okay\_to\_reduce\_gradT\_excess$). This is the option that makes the small stars in the mass range of  $10^5M_\odot \le  M < 9 \times 10^5 M_\odot$. Even with this energy transport boosting the small stars are fully convective. We present the density profiles of two models of the same mass but with different radii in Fig. \ref{fig:rho_vs_R_R9_10}. Model 10 has the energy transport boosting. Practically both model are (almost) fully convective.  }}}}

\subsection{Non-default parameters identical to all runs}
\label{subsecAdefault}

Below we list parameters that we changed from their default values in the same manner in all runs. 
 \begin{itemize}

\item Opacity. We used the default setting for the opacity tables ($kappa\_file\_prefix = \it{'gs98'}$). We activated $use\_Type2\_opacities$ and took `Zbase' to be identical to the metallicity which we set at 0.

\item Mixing parameters.
The parameter $mixing\_length\_alpha$ times a local pressure scale height is the mixing length. The default of \textsc{MESA} is $mixing\_length\_alpha = 2$. For all of the simulations presented in Table 1 we used $mixing\_length\_alpha = 1.5$ \citep{ShiodeQuataert2014, Fulleretal2015}. The parameter $alpha\_semiconvection$ which determines efficiency of semiconvective mixing is taken as 1 (default is 0). We took $num\_cells\_for\_smooth\_gradL\_composition\_term$ and $threshold\_for\_gradL\_composition\_term$ as 10 and 0.02, respectively, to help with convergence. The parameter ${use\_Ledoux\_criterion}$ was activated since thermohaline mixing and semiconvection only applies when $use\_Ledoux\_criterion$ is activated.
Regarding the thermohaline coefficients, we used the Kippenhahn method (\citealt{Kippenhahnetal1980}, default option of \textsc{MESA}) for the $thermohaline\_option$ parameter, while for the $thermohaline\_coeff$ parameter which determines efficiency of thermohaline mixing  we took 1 (the default in \textsc{MESA} is 0). 
  
\item Rotation. We checked rotation for fast rotators. We take the angular velocity $\Omega$ as $0.3\Omega_c$ where $\Omega_c$ is the critical angular velocity, $\Omega_c=\sqrt{\left(1-\frac{L}{L_{Edd}}\right)\frac{GM}{R^3}}$, L, M and R are the luminosity, total mass and photospheric radius of the star, respectively, and $L_{Edd}$ is the Eddington luminosity (for more details and relevant references see \citealt{Paxtonetal2013, Gofmanetal2018}). For slow rotating SMSs we were not able to achieve convergence at these high mass stars, however not all set of parameters were tried. 

\item Atmosphere. We set the atmosphere as the default  option of \textsc{MESA} $simple\_atmosphere$. However, we increased the $Pextra\_factor$ (which is the parameter in-charge for extra pressure in surface boundary conditions) to be $2$, to help with convergence (as the manual of \textsc{MESA} recommends).
   
\item Mass-change. {{{{We followed the prescription of mass-loss from \cite{OuchiMaeda2019}}}}}. 
From different runs we conducted and existing examples found in \textsc{MESA}, it is our understanding that \textsc{MESA} does not allow high accretion rates for such massive stars, and hence we do not include accretion. We made small changes to the three parameters of $max\_logT\_for\_k\_below\_const\_q$, $max\_logT\_for\_k\_const\_mass$, and 
$min\_q\_for\_k\_const\_mass$, from their default values of  $1$, $1$ and $0.99$ to the values of $0.99$, $0.98$ and $0.98$, respectively.

{{{{We note that the mass-loss increases with stellar radius. Smaller stars have much lower mass-loss rates. The $10^6M_\odot$ models, for example, lose only $<0.1\%$ of their mass during our simulation because they are short-lived and they have smaller radii than about the other half of the stars. Stars on the lower radii range lose even less mass. This is the reason that for these stars the final mass after rounding in Table \ref{tab:my-table} is as their initial mass.}}}}

\item Time-step controls. We set the minimum time-step to be $min\_timestep\_limit=10^{-12}\s$, and took $varcontrol\_target = 10^{-5}$.
The $varcontrol\_target$ parameter is the target value for relative variation in the structure from one model to the next.

\item Stopping condition. For the pre-main sequence (PMS) phase we used the routine of \cite{Shiode2013} and \cite{ShiodeQuataert2014}. We stop our post-main sequence stellar evolution when the central mass fraction of hydrogen ($^1$H) drops below $10^{-5}$. 

In order to limit the total time of the run we chose on several runs to limit $max\_model\_number$ as indicated in the table for each run. The mark `\rm{!}' in the last column of Table \ref{tab:my-table} means that we set no limit on the number of models, i.e., we worked with the default options. As can be seen from comparing the final radius of runs 25 with 17, or 21 with 22, the limit on this parameter does not affect much the final result.

\item Mesh adjustment. The mesh coefficients were kept according to \textsc{MESA} defaults, except for $max\_allowed\_nz$ that was changed throughout all of our runs to $10^5$.

\end{itemize} 
\end{document}